\documentclass[twocolumn,superscriptaddress,showpacs,pra]{revtex4-1}

\usepackage[pdftex]{graphicx}

\begin{document}
\title{Lattice modulation spectroscopy of one-dimensional quantum gases: \\Universal scaling of the absorbed energy}

\author{R. Citro}
\affiliation{Dipartimento di Fisica "E.R. Caianiello", Universit\`a
  degli Studi di Salerno and CNR-SPIN c/o University of Salerno, Via Giovanni Paolo II, 132, I-84084 Fisciano
  (Sa), Italy}
\affiliation{INFN, Sezione di Napoli, Gruppo collegato di Salerno, I-84084 Fisciano (SA), Italy}
\author{E. Demler}
\affiliation{Department of Physics, Harvard University, USA}
\author{T. Giamarchi}
\affiliation{DQMP, University of Geneva, 24 Quai Ernest-Ansermet, CH-1211 Geneva, Switzerland}
\author{M. Knap}
\affiliation{Department of Physics and Institute for Advanced Study, Technical University of Munich, 85748 Garching, Germany}
\affiliation{Munich Center for Quantum Science and Technology (MCQST), Schellingstr. 4, D-80799 M{\"u}nchen, Germany}
\author{E. Orignac}
\affiliation{University of Lyon, Ens de Lyon, University Claude Bernard, CNRS, Laboratoire de Physique, F-69342 Lyon, France}

\begin{abstract}
Lattice modulation spectroscopy is a powerful tool for probing
low-energy excitations of interacting many-body systems. By means of bosonization we analyze the absorbed power in a one dimensional interacting quantum gas of bosons or fermions, subjected to a periodic drive of the optical lattice.
For these Tomonaga Luttinger liquids we find a universal $\omega^3$ scaling of the absorbed power, that at very low-frequency turns into an $\omega^2$ scaling when scattering processes at the boundary of the system are taken into account. We confirm this behavior numerically by simulations based on time-dependent matrix product states.
Furthermore, in the presence of impurities, the theory predicts an $\omega^2$ bulk scaling. While typical response functions of Tomonaga Luttinger liquids are characterized by exponents
that depend on the interaction strength, modulation spectroscopy of cold atoms leads to a universal powerlaw exponent of the absorbed power. Our findings can be readily demonstrated in ultracold atoms in optical lattices with current experimental technology.
\end{abstract}

%\date{\today}
\maketitle

%\tableofcontents

\section{Introduction}

Cold atomic systems offer an unprecedented level of control on the properties of interacting quantum systems \cite{bloch_cold_lattice_review,esslinger_annrev_2010},
and allow  for the realization of a plethora of novel phases and phenomena that were previously inaccessible in other experiments. They have given access to the ``experimental solution'' of certain models, and hence can be referred to as quantum simulators. Among those phenomena, one paradigmatic example for bosons on a lattice is the transition between a superfluid and Mott insulator state.
Such a transition was successfully observed in three-dimension (3D) \cite{greiner_mott_bosons_optical}, 2D~\cite{endres_higgs_bosons_shaking}, and in 1D \cite{haller_mott_1d,boeris_mott_1d_cold}.
For the latter, the transition is found to be in the universality class of the Berezinskii-Kosterlitz-Thouless transitions. Cold atoms have thus provided a remarkable way for testing the universal properties of such models.

In order to analyze quantitatively the properties of the correlated phases and the transitions between them, it is important to develop a detailed understanding of different experimental probes. Among them is lattice modulation spectroscopy \cite{stoferle_tonks_optical}.
This technique consists of modulating the amplitude \cite{kollath_bosons_shaking_dmrg, bordia2017, rubioabadal2020} or the phase \cite{tokuno_sieving_conductivity,citro2003} of the optical lattice in which the atoms are trapped. After some time, the energy deposited in the system or the number of doubly occupied states are measured to characterize the underlying state \cite{tokuno2012,greif_magnetism_shortrange_cold}. {In general, driving of the hopping energy provides a novel form of Floquet engineering, which enables atypical Hamiltonians and exotic states of matter to be produced and controlled \cite{pieplow_2018,goldman2019}.}

For bosons this probe can determine energy gaps and thus locate the Mott-to-superfluid transition \cite{stoferle_tonks_optical,haller_mott_1d,boeris_mott_1d_cold}. Moreover, specific modes of the superfluid such as the Higgs mode could be excited,   \cite{endres_higgs_bosons_shaking} which is expected to occur in 3D and 2D superfluids.

Despite this effort, several questions remain to be investigated for bosons and fermions in 1D. For these systems, a symmetry-broken state cannot exist because of strong phase fluctuations even at zero temperature \cite{giamarchi_book_1d}. Hence, only quasi-long range order can exist as characterized by a powerlaw decay of the certain correlation functions. This result is part of the more general properties of Tomonaga-Luttinger liquids (TLL) that are  expected to describe most of the interacting 1D quantum problems \cite{giamarchi_book_1d,cazalilla_review_bosons}.
Given the absence of true long-range superfluid order one may wonder whether the response to shaking, in a one dimensional bosonic system would also show traces of a Higgs mode as   in higher dimensions \cite{endres_higgs_bosons_shaking}. More generally this prompts for an analysis of the response to shaking of a one-dimensional TLL.

In the present paper we perform such an analysis, both in the gapless (superfluid) and gapped phase (Mott insulator). Using a combination of field theoretical and numerical Matrix Product State (MPS) techniques we obtain the
response of the system to the shaking of the optical lattice. We find that this response is a powerlaw of the shaking frequency, with a {\it universal} exponent. This is quite remarkable
in view of the fact that the TLL is normally characterized by responses which show nonuniversal powerlaw behavior, with exponents depending on the
interaction strength. The choice of modulating the amplitude of the optical lattice is important. Would one modulate the phase instead,
the conductivity would be obtained \cite{tokuno_sieving_conductivity} (as periodic phase modulation translates to a periodic force), yielding non-universal exponents, { while the amplitude modulation modifies the tunneling and consequently the kinetic energy of the system.}

Our results also show that a well formed Higgs mode does not exist in one-dimension, as the response to shaking is monotonically increasing
up to shaking frequencies of the order of the bandwidth of the system. In order to connect to experiments, we also analyze and
discuss the effect of this response to open boundary conditions, as realized for example with a box potential, which constitute a relevant
perturbation at low frequencies. In the presence of impurities similar effects are observed, provided that the concentration of impurities is small
and one can simply add energy absorption due to different impurities.

While the focus of this paper is on one dimensional systems, we note that a similar
approach can be used to describe modulation experiments in higher dimensional systems that allow for a hydrodynamic description. Time-periodic modulation of the interaction strength or the transverse confinement potential will result in the resonant parametric generation of excitation pairs~\cite{Staliunas2002,Staliunas2004,Engels2007,Nath2010}, analogous to the 1D case discussed in this paper. The analysis of the absorbed energy can be done following the same approach that we discuss here, with the main difference being the phase space for collective modes. In the summary section we comment on the relevance of our analysis beyond ensembles of cold atoms and suggest possible applications of our results to chains of Josephson junctions and pump and probe experiments in electron systems.

The paper is organized as follows. In Sec.~\ref{sec:model}, we introduce the models that we are investigating. In Sec.~\ref{sec:bosonization} we discuss the analytical
calculation of the absorbed power of a one-dimensional gas
subject to a lattice modulation within a Tomonaga-Luttinger liquid treatment and compare the low-energy behavior with results obtained from time-dependent Matrix Product States. In Sec.~\ref{sec:trans} we discuss the edge effects
treated through an effective boundary potential that couples to the density of the fluid and also analyze the effect of a single impurity in the bulk of the system. In Sec.\ref{sec:gapped}, we consider modulation spectroscopy in the case of gapped systems such as they would occur for the Bose- and the Fermi-Hubbard model with repulsive interactions and commensurate filling.  In Sec.\ref{sec:conclusions}, we present our summary and discuss our findings.

\section{Models}\label{sec:model}

We consider fermionic or bosonic ultracold atoms confined to a 1D tube
subjected  to a deep lattice potential. For deep enough potentials,
such a system can be described in a tight-binding approximation by a
Hubbard model \cite{georges_leshouches_cold_lectures}.  This leads to
the Bose-Hubbard model for spinless bosons
\cite{bloch_cold_lattice_review}
\begin{eqnarray}\label{eq:bose-hubbard}
  H_{\text{b}}^0 = \sum_l \left[-J_0 (b^\dagger_{l+1} b_l + b^\dagger_l b_{l+1}) + \frac{U}{2} n_l
  (n_l-1)  \right],
\end{eqnarray}
where $b_l^\dagger$ ($b_l$) creates (annihilates) a particle on site
$l$ and $n_l=b^\dagger_l b_l$ counts the particles on site $l$, and
the Fermi-Hubbard model \cite{esslinger_annrev_2010} for spin-1/2
fermions:
\begin{eqnarray} \label{eq:fermi-hubbard}
  H_{\text{f}}^0 = \sum_{l,\sigma} \left[-J_0 (c^\dagger_{l+1,\sigma} c_{l,\sigma} +
    c^\dagger_{l,\sigma} c_{l+1,\sigma}) + \frac U 2  n_{l,\sigma}
    n_{l,-\sigma}   \right],
\end{eqnarray}
where again $c_{l,\sigma}^\dagger$ and $c_{l,\sigma}$ are the creation
and annihilation operators, respectively.

In modulation spectroscopy, the trapped atoms are probed by modulating
the strength of the longitudinal periodic potential. The modulation
lowers or raises the potential barrier between two consecutive minima,
and thus to leading order modifies the strength of the tunneling
$J_0$, as well as the interaction within one well. Modulation
of $J$ is expected to be much larger, since it depends exponentially on the barrier height.
We note that
modulation of the Hamiltonian as a whole does not lead to energy absorption and what is important
is the difference in the relative modulation strength of the two terms in the Hamiltonian.
Therefore, we consider the time-dependent Hamiltonian, in which only the tunneling amplitude is modulated
$J_0+\delta J(t) $, giving rise to \begin{equation}
  \label{eq:ham-perturbed}
  H_\nu(t)=H_\nu^0 + \delta J(t) O_\nu,
\end{equation}
with $\nu\in\mathrm{b,f}$ and:
\begin{eqnarray}
  \label{eq:o-bosons}
  \mathcal{O}_{\text{b}}=\sum_l (b^\dagger_{l+1} b_l + b^\dagger_l b_{l+1}),
\end{eqnarray}
for bosons and
\begin{eqnarray}
  \label{eq:o-fermions}
  \mathcal{O}_{\text{f}}=\sum_{l,\sigma} (c^\dagger_{l+1,\sigma} c_{l,\sigma} +
    c^\dagger_{l,\sigma} c_{l+1,\sigma})
\end{eqnarray}
for fermions. The label $b,f$ refers to fermions and bosons,
respectively.

We work in the linear response limit, with
$\delta J(t) =\delta J \cos (\omega t)$ and $\delta J \ll J_0, U$.  In
linear response, the absorbed power is given by \cite{landau_statmech}
\begin{eqnarray} \label{eq:absoprtion}
  \mathcal{P}=\frac \omega 2 |\delta J|^2 \mathrm{Im}\chi_\nu(\omega),
\end{eqnarray}
where
\begin{eqnarray}
  \label{eq:chi-definition}
  \chi_\nu(\omega)=i \int_0^{+\infty} e^{i\omega t}  \langle
  [\mathcal{O}_\nu(t),\mathcal{O}_\nu(0)]\rangle,
\end{eqnarray}
is the response function. In Sec.~\ref{sec:bosonization}, we calculate
this response function in a low-energy/long wavelength limit.

\section{Bosonization} \label{sec:bosonization}

In the low energy/long wavelength limit, interacting bosons and
fermions can be described within an effective continuum theory called
the Tomonaga-Luttinger
liquid \cite{haldane_bosons,emery_revue_1d,solyom_revue_1d,fukuyama_1d_review,schulz_houches_revue,voit_bosonization_revue,schonhammer03_review,gogolin_1dbook,giamarchi_book_1d,nagaosa_book_sc,cazalilla_review_bosons}.
The low energy excitations are phononic collective excitations with a
linear dispersion, that describe density fluctuations (and when
applicable spin fluctuations). In general, spin and density
fluctuations propagate with different velocities, a phenomenon known
as spin-charge separation~\cite{jompol2009,salomon2019}. In that low
energy limit, the original particles appear as coherent states of the
collective modes. As a result, all the observables of the original system
are expressible in terms of the collective modes.  We will thus use in
this section the bosonized representation to calculate the response
function~(\ref{eq:chi-definition}) and hence the resulting absorbed
power.

\subsection{Bosonized representation for fermions}
\label{sec:fermions}

In the fermionic case, away from commensurate filling,
Hamiltonian~(\ref{eq:fermi-hubbard}) has the bosonized
representation~\cite{schulz_houches_revue,giamarchi_book_1d}:
\begin{eqnarray}
  \label{eq:bosonized-fermions}
H_f&=&H_\rho+H_\sigma, \\
H_\rho&=&\int \frac{dx}{2\pi} \left[u_\rho K_\rho (\pi \Pi_\rho)^2 +
  \frac{u_\rho}{ K_\rho} (\partial_x \phi_\rho)^2 \right], \\
H_\sigma&=&\int \frac{dx}{2\pi} \left[u_\sigma K_\sigma (\pi \Pi_\sigma)^2 +
  \frac{u_\sigma}{ K_\sigma} (\partial_x \phi_\sigma)^2 \right] \\ \nonumber
  &-& \frac{2
  g_{1\perp}}{(2\pi\alpha)^2} \int dx \cos \sqrt{8}\phi_\sigma,
\end{eqnarray}
where $u_\rho,u_\sigma$ are respectively the density and spin
velocity, $K_\rho,K_\sigma$ the density and spin Tomonaga-Luttinger
exponents, { where for small $U$ one has $u_\rho=v_F \sqrt{1+U/(\pi v_F)}$,   $u_\sigma=v_F \sqrt{1-U/(\pi v_F)}$,
$K_\rho=\frac{1}{\sqrt{1+U/(\pi v_F)}}$, $K_\sigma=\frac{1}{\sqrt{1-U/(\pi v_F)}}$, $g_{1\perp}=U$ and $[\phi_{\nu}(x),\Pi_{\nu'}(x')]=i\delta(x-x')
\delta_{\nu,\nu'}$. A general expressions can be obtained for arbitrary $U$ as discussed in \cite{schulz_houches_revue,giamarchi_book_1d}.}
For repulsive interactions,
$H_\sigma$ is renormalized to a fixed point
Hamiltonian $H_\sigma^*$ with $K_\sigma^*=1$ and $g_{1\perp}^*=0$,
yielding gapless excitations with linear dispersion
$\omega=u_\sigma |k|$. For attractive interactions without external
magnetic field  the spin
Hamiltonian $H_\sigma$ is gapped
while the density Hamiltonian $H_\rho$
remains gapless \cite{giamarchi_book_1d}.
At half-filling, Umklapp processes are present \cite{giamarchi_book_1d}. They
contribute to the bosonized Hamiltonian (\ref{eq:bosonized-fermions})  a term
\begin{eqnarray}
H_{\mathrm{Umk.}}=-\frac{4 g_3}{(2\pi \alpha)^2}  \int dx  \cos
  \sqrt{8} n \phi_\rho,
\end{eqnarray}
but, as shown in
App.~\ref{app:irr_pert}, when Umklapp processes are irrelevant in the
renormalization group sense~\cite{shankar_spinless_conductivite}, they
add only subdominant contribution to the absorbed power at low
frequency. When the Umklapp processes are relevant, they open a gap in
the spectrum.

In the perturbative limit, since  $\mathcal{O}_{\text{f}}$ is
proportional to the kinetic energy in Eq.~(\ref{eq:fermi-hubbard}),
its bosonized form is simply the bosonized Hamiltonian of
non-interacting spinless fermions divided by $J_0$.
Changing to the spin/charge fields, we
find~\cite{schulz_houches_revue,giamarchi_book_1d}
\begin{eqnarray}
  \label{eq:bosonized-o-fermions}
  \mathcal{O}_{\text{f}}&=& \sum_{\nu=\rho,\sigma}\mathcal{O}_{\mathrm{f},\nu}=\mathcal{O}_{\mathrm{f},\rho}+\mathcal{O}_{\mathrm{f},\sigma}, \nonumber\\
\mathcal{O}_{\mathrm{f},\nu}&=&2 a \sin (k_F a) \int \frac{dx}{2\pi}
  \left[(\pi \Pi_\nu)^2 + (\partial_x \phi_\nu)^2\right].
\end{eqnarray}

Due to spin-charge separation, the absorbed power is
the sum of a spin and a density contribution.
To find an expression of $\mathcal{O}_\mathrm{f}$ applicable away from
the perturbative limit, we note that  $\mathcal{O}_\mathrm{f}$ is
obtained by differentiating the Fermi-Hubbard
Hamiltonian~(\ref{eq:fermi-hubbard}) with respect to $J_0$. Assuming
that the identity carries over to the bosonized description, we
have
\begin{eqnarray}
  \label{eq:ofrho-nonpert}
  \mathcal{O}_{\text{f},\rho}=\int \left[\frac{\partial (u_\rho K_\rho)}{\partial
  J_0}(\pi \Pi_\rho)^2 + \frac{\partial}{\partial
  J_0}\left(\frac{u_\rho} {K_\rho}\right) (\partial_x \phi_\rho)^2 \right] \frac{dx}{2\pi},\nonumber \\
\end{eqnarray}
and a similar approximation for $\mathcal{O}_{\text{f},\sigma}$. As a
further approximation, in the repulsive case,
we take the fixed point values in $H_\sigma$,
and write
\begin{eqnarray}
  \label{eq:ofsigma-nonpert}
  \mathcal{O}_{\text{f},\sigma}=\int \frac{\partial u_\sigma}{\partial
  J_0} \left[(\pi \Pi_\sigma)^2 + (\partial_x \phi_\sigma)^2
  \right] \frac{dx}{2\pi}.
\end{eqnarray}
Applying Eqs.~(\ref{eq:ofrho-nonpert})--(\ref{eq:ofsigma-nonpert}) in
the perturbative case, Eq.~(\ref{eq:bosonized-o-fermions}) is
recovered.
It is important to note that the full expression of the fermion
kinetic energy contains besides the linear dispersion valid near the
Fermi points corrections coming from band curvature. So the
expressions~(\ref{eq:bosonized-fermions}), (\ref{eq:ofrho-nonpert})
and (\ref{eq:ofsigma-nonpert}) are really the most relevant terms in
an expansion of the operator $\mathcal{O}_{\text{f}}$ in a series of
operators of increasing scaling dimensions. The contributions of
operators of higher scaling dimensions are subdominant at low
frequency as shown in App.~\ref{app:irr_pert}.

\subsection{Bosonized representation for bosons}
\label{sec:boso-boso}

In this Section we turn to bosons. The Bose-Hubbard Hamiltonian with $U>0$  has the bosonized
representation
\begin{eqnarray}
  \label{eq:ham}
  H=\int \frac{dx}{2\pi} \left[u K (\pi \Pi)^2 + \frac u K (\partial_x\phi)^2 \right],
\end{eqnarray}
where $\Pi(x)$ and $\phi(x)$ are conjugate operators that describe the boson density fluctuations,
$u$ is their velocity, and $K$ the Tomonaga-Luttinger
exponent \cite{haldane_bosons} { that to lowest order approximation are determined by
$uK=\frac{\pi \rho_0}{m}$ and $\frac{u}{K}=\frac{U}{\pi}$, where $\rho_0$ is the boson density, while their dependence on general values of the interaction can be found in \cite{cazalilla_correlations_1d,cazalilla_review_bosons,rachel2012,laeuchli2013}. In App. \ref{app:LL-parameters} the Luttinger parameters of the Bose-Hubbard model as a function of the interaction $U$ and system size $L$ are shown.}
At integer filling, Umklapp processes contribute a term $\propto \cos 2 \phi$
to the Hamiltonian~(\ref{eq:ham}). Similarly to the fermionic case,
their contribution is subdominant as long as the system remains in a
Tomonaga-Luttinger liquid ground state.
The
limit $U\to 0$ of the Hamiltonian~(\ref{eq:ham}) is singular, with the
velocity $u$ vanishing to recover the quadratic dispersion of
non-interacting bosons above a condensate, and the Tomonaga-Luttinger
exponent going to $+\infty$.
Thus, in contrast to the fermionic case of Sec.~\ref{sec:fermions}, it
is impossible to
derive a bosonized representation of (\ref{eq:o-bosons}) by
considering the non-interacting limit. However, assuming as in
Sec.~\ref{sec:fermions} that the identity
$\mathcal{O}_{\mathrm{b}}=\frac{\partial H_{\mathrm{b}}}{\partial J_0}$
is applicable to the bosonized Hamiltonian~(\ref{eq:ham}), we find
\begin{eqnarray}
  \label{eq:modulation-b}
  \mathcal{O}_b=\int \frac{dx}{2\pi} \left[ \frac{\partial (u K)}{\partial J_0} (\pi \Pi)^2 +  \frac{\partial}{\partial J_0}\left(\frac{u}{K}\right)
    (\partial_x\phi)^2 \right].
\end{eqnarray}
This expression is similar
to~(\ref{eq:bosonized-o-fermions}). Moreover, in the hard core limit
$U\to +\infty$, bosons can be mapped to non-interacting spinless
fermions~\cite{jordan_transformation}, and  the fermionic
expression~(\ref{eq:bosonized-o-fermions}) yields an explicit
form of $\mathcal{O}_b$ which fully agrees
with~(\ref{eq:modulation-b}). As we discussed in the fermionic case,
the expression~(\ref{eq:modulation-b}) is only the first term in a
series of operators of increasing scaling dimension that represent the
various band curvature terms coming from the dispersion of the lattice
model.

\subsection{Response function in an infinite system}\label{eq:thermo}

With repulsive interactions, both for fermions and for bosons, the calculation of the response
function~(\ref{eq:chi-definition}) reduces to
the calculation of the response function of an operator of the
form $\int dx [A \Pi^2 + B (\partial_x \phi)^2]$ for a Hamiltonian
quadratic in $\Pi$ and $\partial_x \phi$. That calculation is further
simplified by rewriting the bosonized form of the operator
$\mathcal{O}_{\mathrm{b,f}}$ as linear combination of the Hamiltonian
and an operator proportional to $\int (\partial_x \phi)^2$. In the
bosonic case,
\begin{equation}
  \label{eq:o-b-simp}
  \mathcal{O}_{\mathrm{b}}=\frac{1}{u K} \frac{\partial(uK)}{\partial J_0} H -\int
  \frac{dx}{\pi} \frac{u}{K^2} \frac{\partial K}{\partial J_0} (\partial_x \phi)^2,
\end{equation}
 and in the fermionic case, for the perturbative limit,
 \begin{eqnarray}\label{eq:o-f-simp-pert}
   \mathcal{O}_{\mathrm{f},\nu}=2 a \sin(k_F a) \left[\frac{H_\nu}{u_\nu K_\nu} + \left(1-\frac 1
     {K_\nu^2} \right) \int dx (\partial_x \phi_\nu)^2\right],\nonumber \\
 \end{eqnarray}
 while in the non-perturbative limit,
 \begin{eqnarray}\label{eq:o-f-simp}
   \mathcal{O}_{\mathrm{f},\nu}= \frac{\partial}{\partial J_0} (u_\nu
   K_\nu) \left[\frac{H_\nu}{u_\nu
   K_\nu}  - \int  \frac{u_\nu}{K_\nu^2} \frac{\partial K_\nu}{\partial J_0} (\partial_x \phi_\nu)^2\right],\nonumber \\
 \end{eqnarray}
The Hamiltonian being time independent, the response function reduces
up to a
proportionality factor to the one of $\int dx (\partial_x \phi)^2$.
We note that this is the same response function as in the case where
the on-site interaction is modulated.
Furthermore, according to Eq.~(\ref{eq:o-f-simp}),
the response function~(\ref{eq:chi-definition}) vanishes for
non-interacting fermions since $K_\rho=K_\sigma=1$ for any $J_0$ in
that case. This can be established
more directly from the lattice Hamiltonian by noting that for $U=0$,
$\mathcal{O}_{\text{f}}$ is proportional to the Hamiltonian.
More importantly, Eq.~(\ref{eq:o-f-simp})
also shows that the contribution of the spin excitations calculated at
the fixed point $K_\sigma^*=1$ is vanishing.  This indicates that for
interacting fermions the dominant contribution comes from the density
response. Due to the fact that the drive is
coupling only to the density and not to the spin, this is
expected to be the case on general grounds.  Similarly,
in the bosonic case, in the limit $U\to +\infty$, where $K=1$
independent of $J$, the response function~(\ref{eq:chi-definition}) is also
vanishing. Again, this is more directly established by noting that
$\mathcal{O}_\mathrm{b}$ is directly proportional to the hard core
boson Hamiltonian in that limit.

\begin{figure}
	\centering
	\includegraphics[width=9cm]{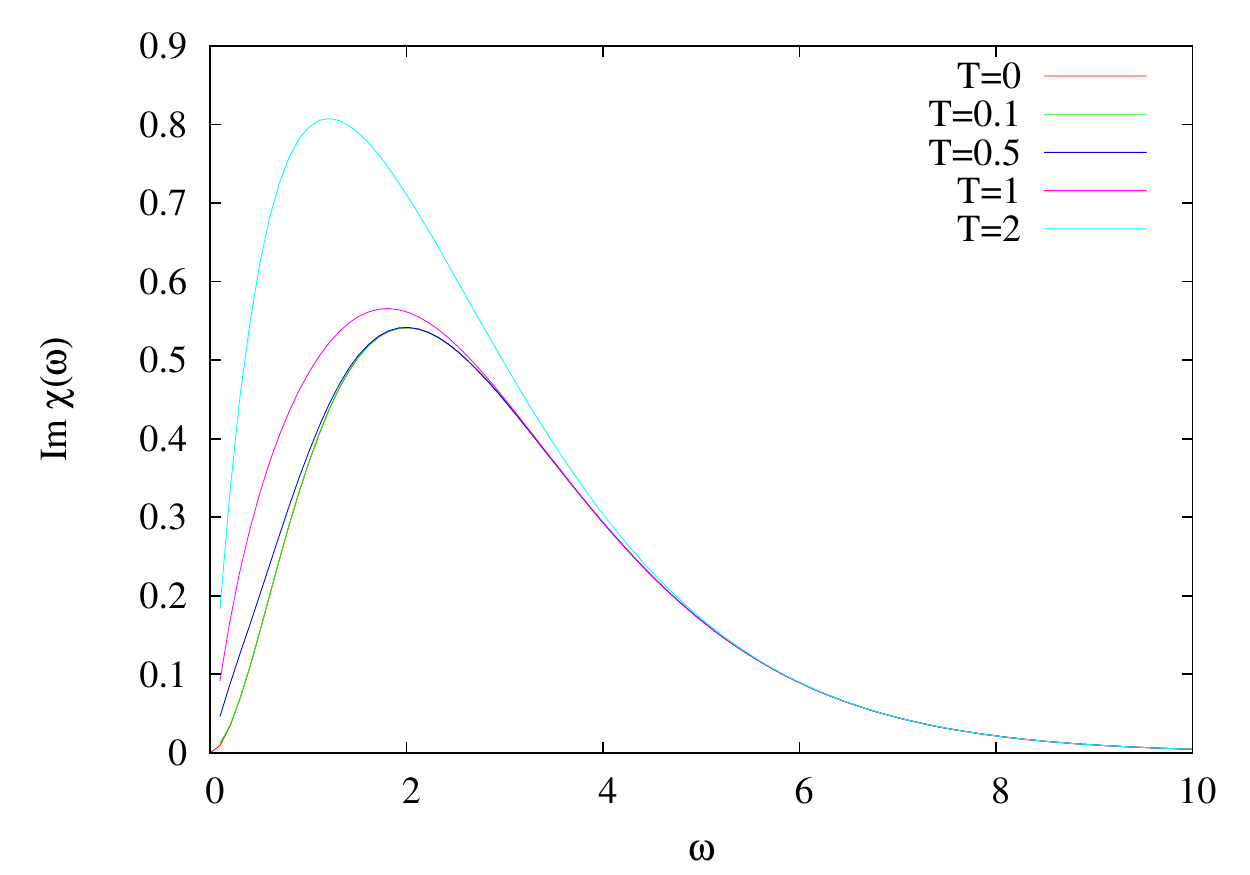}
	\vspace*{0.5cm}
	\caption{Response function $\chi(\omega)$ to a modulation of the lattice at the frequency $\omega$ and temperatute $T$ evaluated from Eq.
 \ref{eq:imchi-finite-t} for $F(K)=1$ and $\alpha/u = 1$.}
	\label{fig:resp}
\end{figure}
We calculate $\chi(\omega)$, by taking the analytic continuation $\chi(\omega)=\chi_M(i\omega_n \to
\omega+i0_+)$  of the Matsubara correlation function
\begin{eqnarray}
  \chi_M(i\omega_n) = \int d\tau e^{i\omega_n\tau} \langle T_\tau O_\nu(\tau) O_\nu(0) \rangle.
\end{eqnarray}
For the sake of definiteness, we perform the calculation for bosons.
Using translational invariance, we find that:
\begin{eqnarray}
&&\frac 1 L \chi_M(i\omega_n)=  \\ \nonumber
&& \left(\frac{u}{\pi K^2}\frac{\partial
  K}{\partial J_0}\right)^2  \int dx
   d\tau e^{i\omega_n \tau}
   \langle T_\tau (\partial_x \phi)^2(x,\tau)  (\partial_x
   \phi)^2(0,0) \rangle.
   \label{eq:corrll}
\end{eqnarray}
Details on the evaluation of $\chi(\omega)$ can be found in the
App.~\ref{app:correlation} and for zero temperature
the final result is
\begin{eqnarray}
\label{eq:imchi}
  \frac 1 L \mathrm{Im}\chi(\omega) = \mathfrak{F}(K) \omega^2
  e^{-|\omega|\alpha/u} \mathrm{sign}(\omega),
\end{eqnarray}
where $\alpha$ is a  short
distance  cutoff (of the order of the lattice spacing)  and
$\mathfrak{F}(K)=\frac{1}{16u}\frac{1}{K^2}\left(\frac{\partial
    K}{\partial J_0}\right)^2$. Only the behavior for $|\omega|\ll u/a \sim J$ is reliably predicted by bosonization.
For frequencies of order of the bandwidth, the linearized approximation for the
dispersion certainly breaks down, and high energy excited states not
described by bosonization can contribute as well to the energy
absorption.

At finite temperature, Eq.~(\ref{eq:imchi}) becomes
\begin{eqnarray}
\label{eq:imchi-finite-t}
  \frac{1}{L} \mathrm{Im}\chi(\omega)=\mathfrak{F}(K) \omega^2
   e^{-|\omega|\alpha/u} \coth \left(\frac{\omega}{4T} \right),
\end{eqnarray}
so the response function behaves as $\sim \omega T$ when $\omega \ll
T$ and as $\omega^2$ when $T \ll \omega$, see Fig.~\ref{fig:resp}.
Thus the absorbed power is
\begin{eqnarray}
  \label{eq:absorbed-power-bosons}
  \mathcal{P}_{\mathrm{b}} = \frac{L}{32 u} |\omega|^3 \left(\frac{\delta J} K
  \frac{\partial K}{\partial J_0}\right)^2
  \coth\left(\frac{\omega}{4T} \right) e^{-\frac{|\omega|\alpha}{u}}
\end{eqnarray}
for bosons and
\begin{eqnarray}
  \label{eq:absorbed-power-fermions}
  \mathcal{P}_{\mathrm{f}} = \frac{L}{32 u_\rho} |\omega|^3 \left(\frac{\delta J} {K_\rho}
  \frac{\partial K_\rho}{\partial J_0}\right)^2
  \coth\left(\frac{\omega}{4T} \right) e^{-\frac{|\omega|\alpha}{u_\rho}}
\end{eqnarray}
for fermions.

It  has a universal power-law dependence on
frequency, with an exponent independent of interactions.
This universal behavior has to be contrasted with the conductivity \cite{giamarchi_umklapp_1d,giamarchi_book_1d} where the power-law exponent varies with the Tomonaga-Luttinger parameter, and thus depends on the
microscopic interaction strength. Here, only the prefactor depends on the logarithmic derivative of the Tomonaga-Luttinger parameter with respect to the hopping amplitude. {In the App.~\ref{app:LL-parameters} the dependence of this prefactor on system size and interaction $U$ is reported.}

\subsection{Numerical results}

\begin{figure}
	\centering
	\includegraphics[width=1.\columnwidth]{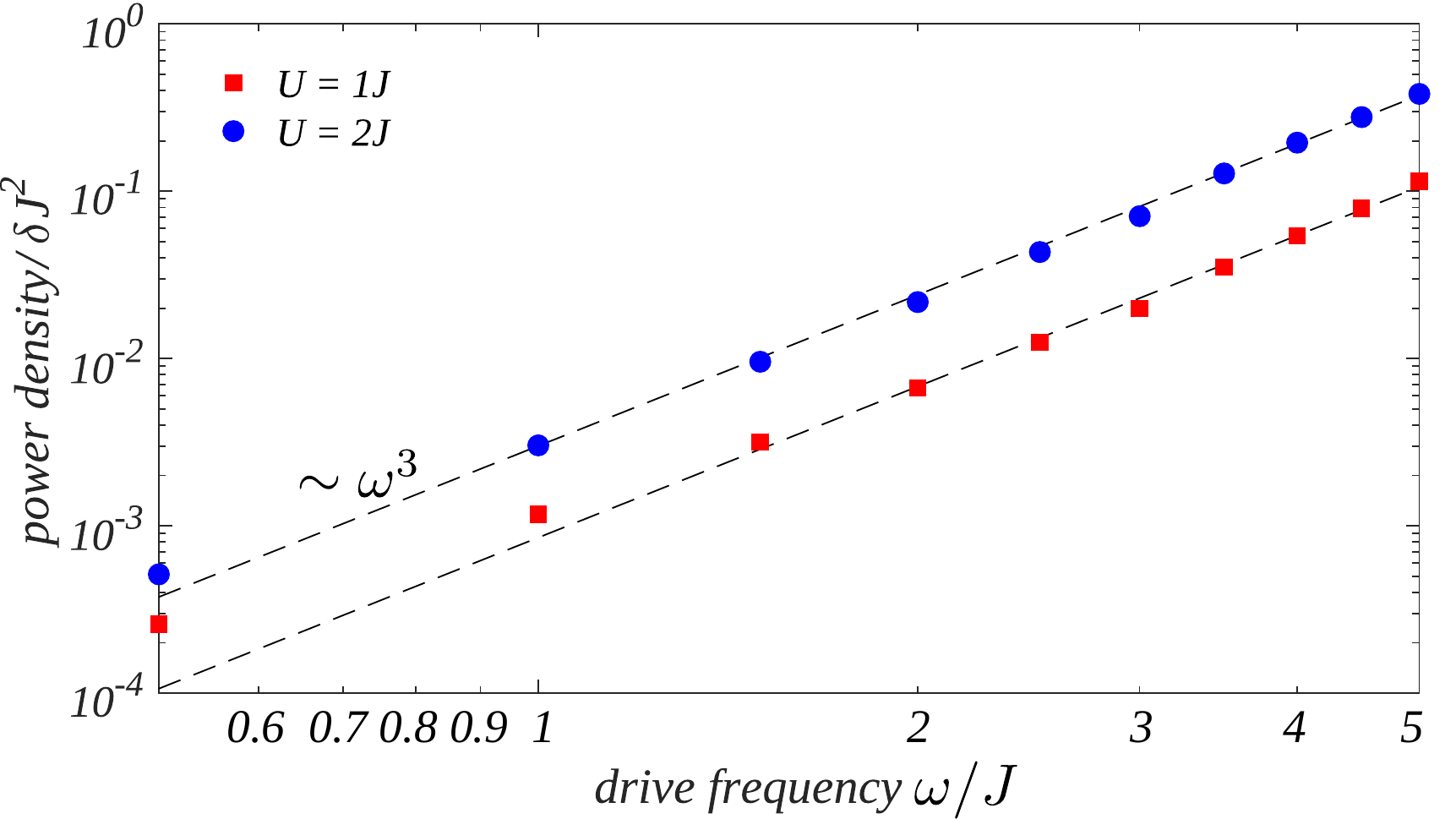}
	\caption{\textbf{Absorbed power density.} Using matrix product states, we have evaluated the absorbed power in a periodically driven Bose-Hubbard model on an open chain of $L=120$ sites and density $\rho = 1.2$ for two values of the interaction strength (see legend). The absorbed power density, renormalized by the drive strength $\delta J^2$, shows a universal $\omega^3$ scaling, as predicted from the Tomonaga Luttinger theory. Deviations at low frequencies are decreasing with increasing system size and are expected to arise from the residual contributions of system edges which add a  $\omega^2/L$ contribution that is leading in frequency but vanishing in the thermodynamic limit.}
	\label{fig:power}
\end{figure}

In order to elucidate the universal frequency exponent of the absorbed power density $\mathcal{P}_b/L \sim \delta J^2 \omega^3$ predicted from the Tomonaga Luttinger theory, we numerically evaluate the energy absorption in the Bose-Hubbard model, Eq. (\ref{eq:bose-hubbard}) using matrix product states~\cite{verstraete2008,schollwoeck2011}. In particular, we consider systems with 120 sites and non-integer boson density $\rho=1.2$, to fully avoid Umklapp processes. We have checked the convergence of our results with the bond dimension of the matrix product state which ranges from $\chi=400$ to $\chi = 800$.

Our objective is to simulate an experimental protocol to measure the absorption. To this end, we first compute the ground state of our model and then apply a periodic modulation of the kinetic energy of the form $J(t) = J_0+\delta J\sin(\omega t)$. We choose the driving strength $\delta J=0.1 J_0$, small enough, such that the absorbed energy increases linearly in time, as required from the linear response theory. We evolve the system for a few drive periods and extract the absorbed power density for a range of modulation frequencies, see Fig.~\ref{fig:power}. The power density scales as $\omega^3$ in agreement with the Tomonaga Luttinger liquid prediction. At low frequencies there are small deviations from the predicted scaling, as expected from a contribution from boundaries (see Sec.~\ref{sec:boundaries}).

\section{Broken translational symmetry}
\label{sec:trans}

So far, our considerations  have been restricted to an infinite
system without defects. This section focuses instead on the case of
systems with broken translational symmetry caused either by boundaries
or impurities.

\subsection{Effects of boundaries}
\label{sec:boundaries}

Since trapped atoms are systems of finite length the effect of boundaries on their response must in principle be considered.
We examine in this section the effect of edge potentials that pin the density and this can potentially modify the response to shaking.

The bosonized Hamiltonian in the presence of forward scattering edge potentials becomes~\cite{eggert_openchains,fabrizio_open_electron_gas,brunel99_edges_logs,affleck_edge_xxz,giamarchi_book_1d}:
\begin{eqnarray}
 H&=&\int_0^L \frac{dx}{2\pi} \left[u K (\pi \Pi)^2 + \frac u K (\partial_x\phi)^2 \right] \nonumber \\
&-& \frac{V}{\pi} \left[\partial_x \phi(0) +\partial_x \phi(L)\right]
\label{eq:ham_ob}
\end{eqnarray}
with the Dirichlet boundary conditions $\phi(0)=0$ and $\phi(L)=-\pi
N$.
Those boundary
conditions ensure that no current can leak through the edges of the
system. As $\rho(x)=-\partial_x \phi/\pi$, the terms $V$ simply
represent a forward scattering in the vicinity of the system
edges. Note that with the Dirichlet boundary conditions, a
backscattering term $-V_b \cos 2\phi(0)$ can be reduced to a forward
scattering term~\cite{giamarchi_book_1d} so there is no loss of
generality in Eq.~(\ref{eq:ham_ob}).
Since in bosonization, the particle-hole symmetry is $\phi \to -\phi$
and $\Pi \to -\Pi$, $V$  vanishes in a particle-hole
symmetric system~\cite{brunel99_edges_logs}.
In the absence of such symmetry however, those terms
can  be nonzero.
When one considers only the static properties, the edge potential
can be eliminated by modifying the Dirichlet boundary
conditions~\cite{affleck_edge_xxz}.
However, when we modulate the lattice, the edge potential can be time
dependent $V=V(J(t))$, and for that reason, it is better to retain the original
boundary conditions.
When we differentiate the Hamiltonian~(\ref{eq:ham_ob})  with respect
to $J_0$, as in Eq.~(\ref{eq:modulation-b}), the edge potential in
~(\ref{eq:ham_ob})
gives an extra edge contribution proportional to $\partial_{J_0} V$ to
the operator $\mathcal{O}_{\text{b}}$
\begin{eqnarray}
&& O_b=\int_0^L \frac{dx}{2\pi} \left[\frac{\partial}{\partial J_0}
  (uK) (\pi \Pi)^2 + \frac{\partial}{\partial J_0}
  \left(\frac u K \right) (\partial_x \phi)^2\right] \nonumber \\
&& - \frac 1 \pi \frac{\partial V}{\partial J_0} [\partial_x\phi(0) + \partial_x
  \phi(L)].
\label{eq:o-b-edges}
\end{eqnarray}
The response coming from the edge potential is calculated in
App.\ref{app:boundaries}. It contributes
\begin{eqnarray}
  \label{eq:edge-power}
  \mathcal{P}_{edge}=\left(\frac V u\right)^2 \left[\delta J
  \frac{\partial}{\partial J_0}\left(\ln \frac{VK}{u}\right)\right]^2
  \frac{\omega^2}{\pi},
\end{eqnarray}
to the absorbed power.
The total absorbed power
is therefore $P_{\mathrm{tot.}}=P_{\mathrm{edge}} + P_{\mathrm{bulk}}
\sim \omega^2 + L \omega^3$. The edge response dominates below a
crossover frequency $\omega^* \sim 1/L$. The boundary potential $V$
remains to be determined. A possible approach is to see how the Friedel
oscillations are affected by these scattering potentials at the boundary.

\subsection{Friedel oscillations and determination of the edge potentials}

 The edge potential $V$ in Eq.~(\ref{eq:ham_ob}) can be deduced  from
 the Friedel
 oscillations~\cite{friedel_oscillations,egger_friedel,rommer00_friedel1d}
 in the density profile of the ground state.
 The explicit calculation of the density
 profile in App.~\ref{app:friedel_oscillations} leads us to the
 following expression valid sufficiently far from edges
\begin{equation}\label{eq:friedel-asymp}
\langle\rho(x)\rangle\sim \frac{\cos (2k_F' x -\varphi)}{\left( \sin \left((\frac{\pi x}{L}\right)\right)^K},
\end{equation}
 where $2k_F'=2k_F+ \frac{4 K V}{u L} $ and $\varphi=\frac{2 K V}{u}$,
 with $k_F=\pi N_0/L$ the nominal Fermi wavevector of the Friedel
 oscillations in a system of length $L$ containing $N_0$ bosons.
From (\ref{eq:friedel-asymp}), consecutive zeros of the Friedel
oscillations are separated by the distance
$\frac{\pi}{2k_F'}=\frac{L}{2N_0}-\frac{L}{N_0^2}\frac{KV}{\pi u} + O(L/N_0^3)$, instead
of $\frac{L}{2N_0}$, revealing the presence of the edge potential. In the thermodynamic limit, $2k_F'$ reduces to
$2k_F$. However, the phase shift $\varphi=2KV/u$ persists, and the fitted
expression of the Friedel oscillations  obtained by MPS reveals the presence or absence
of a potential near the edge.

\subsection{Effects of a single impurity}

Let us finally consider a single impurity located at $x_0$ whose potential energy is given by $H_{\text{imp}}=V \rho(x) \delta (x-x_0)$. Within the bosonization approach this term gives rise to two terms in the Luttinger liquid Hamiltonian: a term $-\frac{1}{\pi} \partial_x \phi(x_0)$ which corresponds to a forward scattering process and a term proportional to $\cos (2 \phi(x_0))$ which corresponds to back scattering. Using the same treatment as above, the first term will be leading to a dominant $\omega^2$ scaling function for the absorbed power, while the back scattering will contribute a term proportional to $\omega^{2(K-1)}$ with $K>1$ and thus less relevant at low-frequency. Thus the presence of a single impurity would lead to a dominant $\omega^2$ contribution to the absorbed power.

\section{Gapped systems}\label{sec:gapped}

In the case of fermions with attractive interactions, or in the case
of fermions or bosons with repulsive interactions at commensurate
filling, the spectrum can become gapped. The response in that gapped
regime can be calculated either in the Luther-Emery
limit~\cite{luther_exact,giamarchi_book_1d} or in the more general case
using the form factor
expansion~\cite{essler04_condmat_exact_review}. Both methods predict a
threshold in absorption power at the gap.

For the Bose-Hubbard model, below that gap the power absorption will be zero. In the Fermi-Hubbard case with repulsive interactions,
we have seen that the spin response was
suppressed, so that only the density response contributed.
In the gapped state, the density response also does not contribute at
frequency lower than the gap, making the threshold observable as well.
In the case of the Fermi-Hubbard model with attractive interaction,
since the density modes are gapless,  the response at low frequency
will be the $\sim |\omega|^3$
contribution. The threshold at the gap then appears as a cusp-shaped
rapid increase of absorption.
The physical interpretation of such threshold is quite simple.
At frequencies lower than the binding energy of two fermions of
opposite spins, the pairs of fermions behave as an interacting boson
gas~\cite{solyom_revue_1d}, yielding the $\sim |\omega|^3$ contribution
to the absorbed power. As the frequency becomes comparable to the
binding energy of the pair, another absorption channel from
dissociation of the pairs becomes available, leading to the rapid
increase of absorption.

For concreteness, let's first consider the Fermi-Hubbard model in the Mott insulating phase for the particular case of Luther-Emery limit where $K_\rho = 1/2$. In that limit, the resulting absorbed power is given by (see App.~\ref{app:luther-emery} for a detailed calculation)
\begin{eqnarray}
\label{eq:le-power}
\mathcal{P}_{LE}=\frac{L\Delta^2} u \left(\frac{5a \delta J} {2u} \sin(k_F
a)\right)^2 \sqrt{\left(\frac\omega 2\right)^2 -\Delta^2},
\end{eqnarray}
where $2\Delta <\omega <4\Delta$ leading to a cusp singularity at $\omega=2\Delta$.

This analysis can be extended away from the Luther-Emery point to any value of of the Luttinger parameter by the form factor expansion for the sine-Gordon
model~\cite{karowski_ff,babujian99_ff_sg_1,babujian02_ff_sg_2,essler04_condmat_exact_review},
as detailed in the App.~\ref{app:form_factor} (see also
\cite{iucci_absorption}). The main result for the absorbed power (with
$1/2 < K_\rho <1$) is

\begin{eqnarray}
\mathcal{P}_{FF} = L  \left(\frac{2K_\rho}{K_\rho^2+1}\right)^2
\frac{2M_s^2 |2a \sin (k_F a) \delta J|^2}{\pi u_\rho^3
  \nu^2}\sqrt{\omega^2-4M_s^2},\nonumber \\
\label{eq:ff}
\end{eqnarray}
where we have found a threshold at twice the mass of the soliton $M_s$
and $\nu=K_\rho/(1-K_\rho)$ for the Fermi-Hubbard model. For $K_\rho
<1/2$, besides the threshold behavior~(\ref{eq:ff}), discrete peaks
coming from resonant absorption by soliton-antisoliton bound states
become possible. This behavior could be readily observed in current experiments with cold atoms
in the Mott insulating regime\cite{Haller2010,boeris_mott_1d_cold}.

The same threshold behavior as in Eqs.~(\ref{eq:le-power},\ref{eq:ff})
was also obtained in the opposite limit of a weak
lattice~\cite{iucci_shake_bosons_theory} in which the depth of the
periodic potential was modulated. One may thus speculate whether
such threshold behavior is also observed for intermediate lattice
strengths.

\section{Summary and Outlook }
\label{sec:conclusions}

We have analyzed in linear response
the power absorbed by one-dimensional fermions
and bosons in the  Tomonaga-Luttinger liquid~\cite{haldane_bosons} or
Luther-Emery liquid~\cite{luther_exact}
phase, to the
amplitude modulation of an optical lattice. In the Tomonaga-Luttinger
liquid, we have found that the
absorbed power possesses a {\it universal} $\omega^3$ power law onset,
that has been confirmed by numerical
simulations based on Matrix Product States.
We have also shown that this power law crosses over to $\sim
\omega^2$, at low frequency in finite systems  when edge effects are taken into
account. A similar $\omega^2$ behavior is found for systems with a
single impurity located in the bulk.

Such universal behavior is surprising since in
Tomonaga-Luttinger liquids  theory, response functions usually show
nonuniversal exponents determined by the interaction
strength~\cite{kollathGiamarchi2006}. The universal
$\omega^3$ scaling of the absorbed power can be readily measured for
ultracold atoms in optical lattices confined to one-dimension by
measuring the energy change over time.
In Luther-Emery liquid phases, that can be obtained for
commensurate densities, or with spin-1/2 fermions having attractive
interaction, the absorbed power vanishes below a gap and shows a
marked onset above, thus making, if possible, to identify this energy scale.

The discussion in this paper focused on experiments with spinless ultracold atoms. Before concluding this section we briefly review other systems in which ideas developed in this paper can be tested experimentally.

Bosonic spin mixtures in optical lattices can be used to realize lattice spin Hamiltonians and spinor condensates~\cite{Duan2003,Widera2008,Stamper-Kurn2013}. Recent experiments by Jepsen et al \cite{Jepsen2010}
used magnetic field dependence of the interspecies scattering length to realize XXZ spin chains with tunable anisotropy of  interactions. In the regime of easy plane anisotropy XXZ chains are in the gapless regime, while the easy axis case corresponds to the gapped regime. Periodic modulation of $J_z/J_\perp$ can be  achieved in this system through periodic modulation of the magnetic field and should have an effect equivalent to modulation of the interaction strength for spinless bosons. These experiments have high local resolution, which will allow one to spatially resolve spin patterns induced by modulation of the interaction anisotropy. Hence predictions of our paper for both gapless and gapped regimes can be checked experimentally.

Recent progress in superconducting nanotechnology makes it possible to engineer arrays of
coupled  Josephson junctions whose parameters can be controlled dynamically. Lahteenmaki et al.~\cite{Lahteenmaki2011} have demonstrated a dynamical Casimir effect in a one dimensional chain of Josephson junctions, in which the Josephson energy of the junctions
has been modulated by periodically changing the background magnetic flux. Parametric generation of photons at half the modulation frequency observed in these experiments is the direct analogue of energy absorption in the Luttinger liquid discussed in our paper. Recent experiments by Kuzmin et al \cite{Kuzmin2019} demonstrated the possibility of tuning a chain of Josephson junctions through the superconductor to insulator transition and explored evolution of the collective phase mode across the transition. Hence 1D superconducting metamaterials make it possible to study modulation spectroscopy of 1D systems in both gapless and gapped phases.

Although the focus of this paper has been on one dimensional systems, a similar analysis can be applied to study periodic driving of higher dimensional systems provided that their lower energy excitations allow field theoretical description.
{ Modulation of the kinetic energy of bosons in optical lattices  has been considered in the context of the Higgs mode in systems with broken U(1) symmetry \cite{Podolsky2011,endres2012,Pekker2015}. In the superfluid phase in $d=2$, $3$ close to the critical point the imaginary part of the response function of the operator of kinetic energy develops a broad peak at the energy equal to the Higgs mode frequency and has a universal scaling form proportional to $\omega^{d+1}$ at smaller frequencies. The latter is determined by the process of resonant excitation of pairs of Goldstone modes with opposite momenta mediated by the virtual excitation of the Higgs mode. This process is equivalent to the mechanism of exciting pairs of Luttinger liquid phonons considered in our paper for one dimensional systems. Thus energy absorption rate at low frequencies has a general scaling form $\omega^{d+2}$}.
We also note that our formalism should be useful for analyzing pump and probe experiments in interacting electron systems \cite{Kampfrath2013,Basov2011,Giannetti2016,Basov2017,Cao2018}. Recent experiments by von Hoegen et al. \cite{VonHoegen2019} have observed parametric excitation of Josephson plasmons in YBCO superconductors following resonant excitation of apical oxygen phonons. The microscopic mechanism of phonon-plasmon coupling is modulation of the superfluid density in copper-oxide planes by the phonon induced motion of oxygen atoms. Analogously to what we have discussed in this paper, resonant parametric excitation of plasmon pairs has been a crucial component of experiments by von Hoegen et al. One important difference, however, is that three wave mixing between phonons and plasmons involves two different types of plasmons, the so-called lower and upper Josephson plasmons. The formalism developed in our paper can be extended to the case of parametric instabilities involving different types of collective excitations. We expect that resonant parametric interactions between phonons and collective excitations of many-body electron systems should be an ubquitous phenomenon. Excited phonons can modulate several parameters of electron systems, including effective mass, interactions, and carrier density. Thus pump and probe experiments can be used to achieve parametric driving of a broad range of collective modes, including plasmons in superconductors, spin waves in magnets, and phasons in incommensurate CDW systems.

\begin{acknowledgments}
  This work was supported in part by the Swiss National Science Foundation under Division II.
  MK acknowledges support from the Technical University of Munich - Institute for Advanced Study, funded by the German Excellence Initiative and the European Union FP7 under grant agreement 291763, the Deutsche Forschungsgemeinschaft (DFG, German Research Foundation) under Germany's Excellence Strategy--EXC-2111--390814868, the European Research Council (ERC) under the European Union's Horizon 2020 research and innovation programme (grant agreement No. 851161), from DFG grants No. KN1254/1-1, No. KN1254/1-2, and DFG TRR80 (Project F8).
  ED acknowledges support from
  Harvard-MIT CUA, AFOSR-MURI: Photonic Quantum Matter (award FA95501610323), DARPA DRINQS program (award D18AC00014). RC acknowledges hospitality at DQMP, Geneve (CH) and at Harvard-MIT CUA.  EO acknowledges hospitality at DQMP, Geneve (CH).

\end{acknowledgments}

\appendix

%%%%%%%%%%%%%%%%%%%%%%%%%%%%%%%%%%%%%%%%%
\section{Irrelevant perturbations}
\label{app:irr_pert}

We want now evaluate the corrections to the response function (\ref{eq:corrll}) in the presence of an irrelevant perturbation $H_p=(\frac{g}{2\pi}) \int dx \int d\tau \cos (2 \phi(x,\tau))$.
To do this we apply  second order perturbation theory and thus we need to evaluate the average of the following  time ordered product of operators:

\begin{equation}
\Phi(x,\tau)=<T_\tau \partial^2_x \phi (x,\tau) \cos(2\phi (1))\cos (2\phi(2))\partial^2_x \phi (0,0)>,
\end{equation}
where we used the compact notation $1\equiv (x_1,\tau_1)$ and similarly for 2. This average can be estimated from the correlator
\begin{equation}
<T_\tau e^{i \lambda \partial_x \phi (x,\tau)} e^{i \mu \partial_x \phi (0,0)} \cos(2\phi (1))\cos (2\phi(2))>,
\end{equation}
taking the second derivative with respect to $\lambda$ and $\mu$ in the limit $\lambda,\mu\rightarrow 0$ and keeping in mind the following identity:
\begin{equation}
<T_\tau \Pi_j e^{i q_j f(\phi(x_j,\tau_j))}>=e^{-\sum_{i>j}q_j <T_\tau f(\phi(x_i,\tau_i))f(\phi(x_j,\tau_j))>}
\end{equation}
where $f(\phi(x,\tau))=\partial_x \phi(x,\tau)$ or $\phi(x,\tau)$.
This leads to the following expression:
\begin{widetext}
\begin{eqnarray}
  && \Phi(x,\tau)=64 \int d1 \int d2 e^{-2\langle T_\tau
     [\phi(1)-\phi(2)]^2\rangle}\langle T_\tau \partial_x \phi(x,\tau) \partial_x \phi(0,0)\rangle  \nonumber \\
  &&  \times \langle T_\tau \partial_x \phi(x,\tau) \partial_x
     \phi(0,0)\rangle  \left[ \langle T_\tau \partial_x \phi(x,\tau)
     \phi(1)\rangle \langle T_\tau
     \phi(1) \partial_x\phi(0)\rangle \right. \nonumber \\
  && \left. - \langle T_\tau \partial_x \phi(x,\tau) \phi(1)\rangle
     \langle T_\tau \phi(2) \partial_x\phi(0)\rangle\right]
\label{eq:phi}
\end{eqnarray}
\end{widetext}
This expression correspond to Hartree and Fock diagrams. Explicitly one has:
\begin{equation}
e^{-2<T_\tau [\phi(1)-\phi(2)]^2>}=\left ( \frac{\alpha^2}{(\alpha+u|\tau_1-\tau_2|)^2+(x_1-x_2)^2}\right)^K
\end{equation}
while
\begin{eqnarray}
&&\langle T_\tau \partial_x \phi(x,\tau)\phi(x_1,\tau_1)\rangle =\partial_x
   G(x-x_1,\tau-\tau_1)\nonumber \\
 && =-\frac{K}{2}\frac{(x-x_1)}{(x-x_1)^2+(u|\tau-\tau_1|+\alpha)^2} \\
&&\langle T_\tau \partial_x \phi(0,0)\phi(x_1,\tau_1)\rangle
   =\partial_x
   G(x_1,\tau_1)=\frac{K}{2}\frac{x_1}{x_1^2+(u|\tau_1|+\alpha)^2}\nonumber
\end{eqnarray}
and similarly for the term with $1\rightarrow 2$. In (\ref{eq:phi}) one can factorize the term
\begin{eqnarray}
&& \int dx \int d \tau \left (
  \frac{\alpha^2}{(\alpha+u|\tau|)^2+x^2}\right)^K \\ &&=\int d\tau
                                                         \frac{\alpha^{2K}}{(\alpha+u|\tau|)^{2K-1}}\int\frac{dy}{(1+y^2)^K}=\frac{2\alpha^2}{u}I(K) \nonumber
\end{eqnarray}
which for $K>1$ gives no power-law correction and recognize the following convolution integral
\begin{eqnarray}
&&\int d\tau_1 \int dx_1 \partial_x G(x-x_1,\tau-\tau_1) \partial_x
   G(x_1,\tau_1)\nonumber \\
  && =\int \frac{dq}{2\pi}\int \frac{d\omega}{2\pi} q^2
     |G(q,\omega)|^2 e^{i(qx-\omega \tau)}.
\end{eqnarray}
\begin{widetext}
Thus the Hartree correction of $\Phi(x,\tau)$ reduces to
\begin{eqnarray}
&& \Phi_H(x,\tau)=\frac{\alpha^2}{u}I(K) \int \frac{dq}{2\pi}\int
   \frac{d\omega}{2\pi} q^2 |G(q,\omega)|^2 e^{i(qx-\omega
   \tau)} \partial_x^2 G(x,\tau)\nonumber \\
&&   =\frac{\alpha^2}{u}I(K) \int \frac{dq}{2\pi}\int \frac{d\nu}{2\pi} q^4 G(q,\nu)G(-q,-\nu)G(q,\omega-\nu),
\end{eqnarray}
\end{widetext}
where $G(q,\nu)=\frac{uK}{\nu^2+(uq)^2}$. One can easily show that
this integral does not increase with $\omega$ so the Hartree correction can be neglected.

The Fock correction instead reduces to the integral
\begin{eqnarray}
\label{eq:fock}
&&\int d1 \int d2 \left (
  \frac{\alpha^2}{(\alpha+u|\tau_1-\tau_2|)^2+(x_1-x_2)^2}\right)^K\nonumber
\\ &&\partial_x G(x-x_1,\tau-\tau_1) \partial_x G(x_2,\tau_2)\partial_x^2 G(x,\tau).
\end{eqnarray}
Turning to the Fourier transform representation, we find
\begin{widetext}
\begin{eqnarray}
\label{eq:self-en}
&&\int dx \int d\tau \frac{e^{-iqx +i\omega \tau}}{\lbrack x^2+(u|\tau|
  +\alpha)^2\rbrack}  \\
  &&=\left ( \sqrt{(q^2+\frac{\omega^2}{u^2})\alpha^2}\right)^{K-1}\frac{1}{2^{K-1}\Gamma(K+1)}K_{K-1}(\sqrt{(q^2+\frac{\omega^2}{u^2})\alpha^2}),\nonumber
\end{eqnarray}
\end{widetext}
where $K$ is the Bessel function of second kind. It can be expanded for small $q, \omega$ and $K-1$ non-integer: while the zeroth order term is $q,\omega$ independent, the first non analytic correction will be of the order $(q^2+\frac{\omega^2}{u^2})^{K-1}$, which is subdominant compared to $(q^2+\frac{\omega^2}{u^2})$ when $K>1$.
In the integral  (\ref{eq:self-en}) we recognize (up to a
factor $g^2$) the self energy-correction from a diagrammatic point of view  and thus we can write that qualitatively
\begin{equation}
\Sigma(q,\omega)\simeq \Sigma (0,0)+C
(q^2+\frac{\omega^2}{u^2})^{K-1},
\end{equation}
neglecting holomorphic terms of order
$(q^2+\frac{\omega^2}{u^2})\alpha^2$ and higher.
We can thus evaluate the Fock correction (\ref{eq:fock}) which is
\begin{eqnarray}
&&\int \frac{d\nu}{2\pi}\int \frac{d q}{2\pi} q^4 \lbrack
   \Sigma(q,\nu)G^2(q,\nu)G(-q,\omega-\nu)\nonumber \\ && +\Sigma(-q,\omega-\nu)G^2(-q,\omega-\nu)G(q,\nu)\rbrack \nonumber \\
&& =g^2\int \frac{d\nu}{2\pi}\int \frac{d q}{2\pi}q^4
   (q^2+\frac{\nu^2}{u^2}\alpha^2)^{K-1}(q^2+\frac{\nu^2}{u^2})^{-2}\nonumber
   \\ && \times (q^2+\frac{(\omega-\nu)^2}{u^2})^{-1}
\end{eqnarray}
By simple power counting this integral behaves as $\omega^{2(K-1)}$ and for $K>2$, it is subdominant compared to the term $\omega^2$ as $\omega \rightarrow 0$.\\
So, when $K>2$ and the Umklapp scattering is irrelevant, the intensity of the modulation spectroscopy behaves as
\begin{equation}
Im \chi(\omega) \sim \omega^2+g^2 \omega^{2(K-1)}+O(\omega^{2K-2}).
\end{equation}
\begin{figure*}
	\centering
	\includegraphics[width=\textwidth]{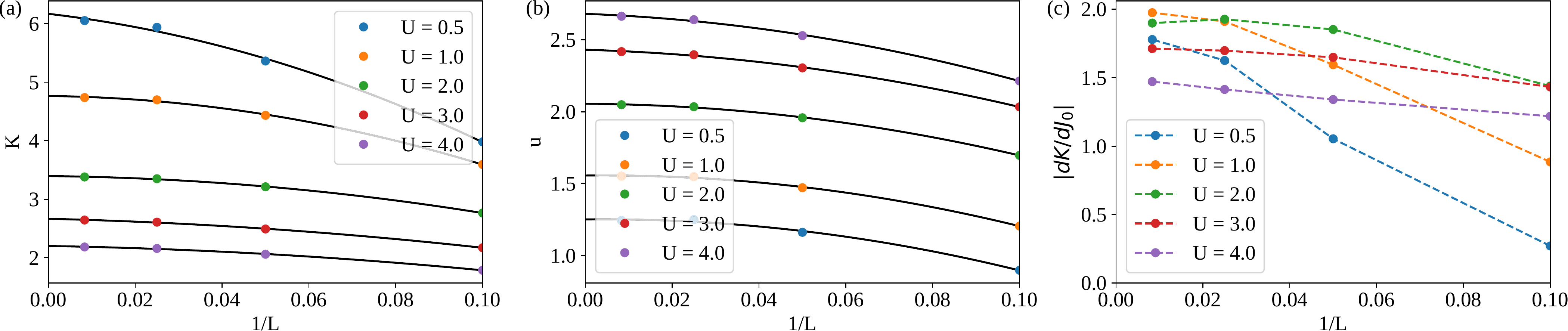}\\
	\caption{\textbf{Luttinger liquid parameters.} We evaluate (a) the Luttinger parameter $K$, (b) the Luttinger velocity $u$, and (c) the derivative of the Luttinger parameter with respect to the kinetic energy $d K/dJ_0$ as a function of the inverse system size $L$ for various values of the interaction strength $U$ and fixed density $\rho_0 = 1.2$. Lines (dashed and solid) are guides to the eyes.}
	\label{fig:LL-parameters}
\end{figure*}

%%%%%%%%%%%%%%%%%%%%%%%%%%%%%%%%%%%%%%%%%%%%%%%%%%%%%%%%%%%%%%%%%%%%%%% 
\section{Luttinger parameters of the Bose-Hubbard model}
\label{app:LL-parameters}

\begin{figure}
	\centering
	\includegraphics[width=1.\columnwidth]{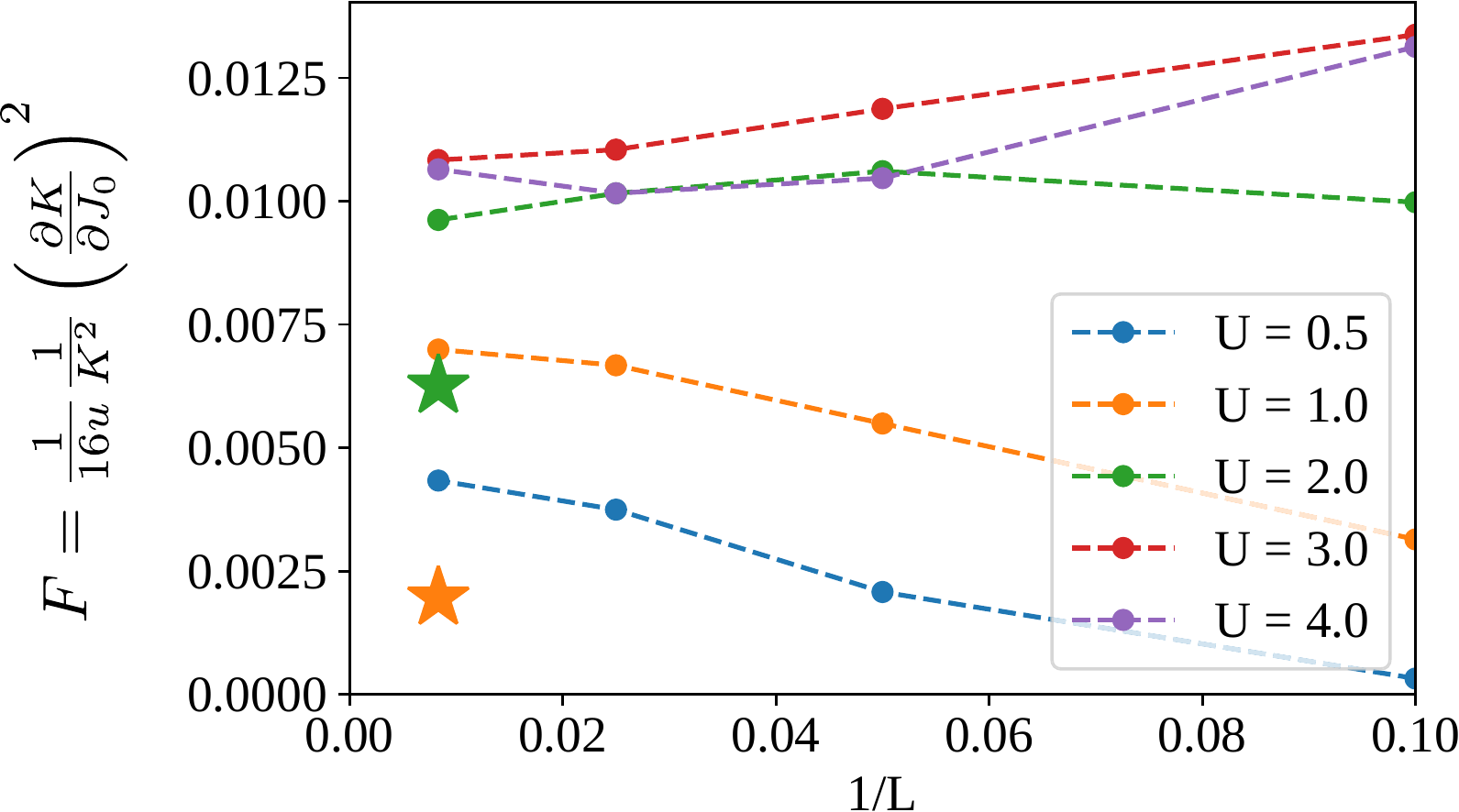}
	\caption{\textbf{Prefactor of the absorbed power.} The analytically predicted prefactor $F(K)$ obtained in Eq. (\ref{eq:imchi}), filled circles, is compared ot the numerically obtained prefactor from the full time evolution, big stars. The data is shown for various values of the interaction strength $U$ and for a fixed density $\rho_0 = 1.2$ as a function of the inverse system size $1/L$.}
	\label{fig:prefactor}
\end{figure}

We evaluate the Luttinger parameters for the Bose-Hubbard model at filling $\rho = 1.2$ using matrix product states. The Luttinger parameter $K$ is obtained from the density-density correlation function and the the Luttinger velocity $u$ from the compressibility $\kappa^-1 = \frac{\partial^2 E}{\partial N^2} = \frac{K}{\pi u}$. The Luttinger parameters are shown as a function of the inverse system size in Fig. \ref{fig:LL-parameters}. From the Luttinger parameters, we can evaluate the prefactor of the absorbed power, as stated in Eq. (\ref{eq:imchi}), see Fig. \ref{fig:prefactor}. We compare the analytically prefactor obtained from Luttinger liquid theory, filled circles, with the prefactor of the absorbed power in the time evolved many-body state, big stars, which is on the same order of magnitude.

%%%%%%%%%%%%%%%%%%%%%%%%%%%%%%%%%%%%%%%%%%%
\section{Evaluation of the retarded correlation function}
\label{app:correlation}
For the sake of definiteness, we present the calculation in the case
of bosons. The fermionic case proceeds along the same line, with a
simple change of prefactor.
Using Wick's theorem, the correlator in Eq.~(\ref{eq:corrll}) is
rewritten
\begin{eqnarray}
   \langle T_\tau (\partial_x \phi)^2(x,\tau)  (\partial_x
   \phi)^2(0,0) \rangle &=& 2  (\langle T_\tau \partial_x \phi(x,\tau)  \partial_x
   \phi(0,0) \rangle)^2 \nonumber \\
 &=& \frac{K^2}{2} \frac{[(u|\tau|+\alpha)^2
   -x^2]^2}{[(u|\tau|+\alpha)^2 +x^2]^4},\nonumber \\
\end{eqnarray}
leading to
\begin{eqnarray}
  \frac 1 L \chi_M(\omega)&=&\frac \pi 8 \left(\frac u {\pi K}
                              \frac{\partial K}{\partial J_0} \right)^2\int_{-\infty}^\infty \frac{e^{i\omega
      \tau}}{(u|\tau|+\alpha)^3} d\tau
\end{eqnarray}
We need the integral
\begin{eqnarray}\label{eq:matsubara-integral}
  I(\omega)=\int_{-\infty}^\infty \frac{e^{i\omega
      \tau}}{(u \tau +\alpha)^3} d\tau,
\end{eqnarray}
and its analytic continuation. We write
\begin{eqnarray}
  \frac{1}{(u |\tau| +\alpha )^3}=\frac{1}{2} \int_0^\infty k^2
  e^{-k (u|\tau|+\alpha)} dk,
\end{eqnarray}
and obtain
\begin{eqnarray}\label{eq:matsubara-int-final}
   I(\omega)=\frac{1}{2} \int_0^{+\infty} dk k^2 e^{-k\alpha}
  \left[\frac {1}{i\omega + uk} - \frac {1}{i\omega - uk}\right].
\end{eqnarray}
We find the analytic continuation $i\omega\to \omega+i {0_+}$  of
Eq.~(\ref{eq:matsubara-int-final}) using
the identity
\begin{eqnarray}
 \frac 1{x+i0_+} = P\left(\frac 1 x\right) -i\pi \delta(x),
\end{eqnarray}
\begin{widetext}
which gives
\begin{eqnarray}
  \label{eq:imag-analytic-cont}
  \mathrm{Im} \chi(i\omega\to \omega+i0_+) &=& \frac L {16}
  \left(\frac u K \frac{\partial K}{\partial J_0} \right)^2 \int_0^{+\infty} dk k^2 e^{-k\alpha}[\delta(\omega-u k) -\delta (\omega+uk)]\\
  &=& \frac L {16 u}
  \left(\frac 1 K \frac{\partial K}{\partial J_0} \right)^2 \omega^2
      e^{-\frac{|\omega|\alpha} u} \mathrm{sign}(\omega).
\end{eqnarray}
This leads to Eq.~(\ref{eq:imchi}) in the main text.
\end{widetext}

%%%%%%%%%%%%%%%%%%%%%%%%%%%%%%%%%%%%%%%%%%%%%%%%%%%%%%%%%

\section{Calculation of the response function in the case of a system
  with boundaries}\label{app:boundaries}

In the case of a system with boundaries described by the
Hamiltonian~(\ref{eq:ham_ob}) with the operator $\mathcal{O}_b$ given
by (\ref{eq:o-b-edges}), we first rewrite
\begin{eqnarray}
  \label{eq:o-b-edge-substr}
  O_b&=&\frac{\partial}{\partial J_0}(uK) \frac{H}{uK} -\int
  \frac{dx}{\pi} \frac u{K^2}  \frac{\partial K}{\partial J_0}
         (\partial_x \phi)^2 \nonumber \\
     && +\frac{V}{\pi} \frac{\partial}{\partial J_0} \left[\ln
        \left(\frac{uK}{V}\right)\right] [\partial_x
        \phi(0)+ \partial_x \phi(L)],
\end{eqnarray}
and as before we only have to calculate the response function of the
bulk term proportional to $(\partial_x \phi)^2$ and the edge term
proportional to $\partial_x \phi(0)+\partial_x \phi(L)$.

To perform the calculation, one first rescales the fields,
$\phi=\sqrt{K} \tilde{\phi}$ and $\Pi=\tilde{\Pi}/\sqrt{K}$ and introduces the
Fourier decomposition (\ref{eq:phi-decomp})--~(\ref{eq:Pi-decomp}) to
rewrite the Hamiltonian~(\ref{eq:ham_ob}) in terms of shifted harmonic
oscillators
\begin{eqnarray}
  \label{eq:ham_ob-harmonic}
  H&=&\frac u 2 \sum_{n=1}^{+\infty} \left[\pi \tilde{\Pi}_n^2+\left(\frac{\pi n}
  L\right)^2 \frac{\tilde{\phi}_n^2}\pi\right] \nonumber \\
&& - \frac V \pi \sqrt{\frac{2K} L}  \sum_{n=1}^{+\infty}
   [1+(-1)^n] \frac{\pi n} L \tilde{\phi}_n,
\end{eqnarray}
and the operator $\mathcal{O}_b$, without the contribution
proportional to the Hamiltonian,
\begin{eqnarray}
  \mathcal{O}_b &=& -\frac{u}{\pi K} \frac{\partial K}{\partial J_0}
  \sum_n \left(\frac{\pi n} L \right)^2 \tilde{\phi}_n^2  \\
  && + \frac{V}{\pi}  \frac{\partial }{\partial J_0} \left[\ln
     \left(\frac{uK}{V}\right)\right] \sqrt{\frac{2K}L} \sum_{n=1}^{+\infty}
   [1+(-1)^n] \frac{\pi n} L \tilde{\phi}_n.\nonumber
\end{eqnarray}
We now introduce $\bar{\phi}_n$ such that
\begin{eqnarray}
  \label{eq:phibar-def}
  \bar{\phi}_n = \tilde{\phi_n} -\frac{L}{\pi n}
  \sqrt{\frac{2K}L}[1+(-)^n] \frac{V}u,
\end{eqnarray}
to have a Hamiltonian purely quadratic in $\bar{\phi}_n$.
In terms of the new operators,
\begin{eqnarray}
  \label{eq:o-b-edge-phibar}
  \mathcal{O}_b&=&-\frac{u}{\pi K} \frac{\partial K}{\partial J_0}  \sum_n
  \left(\frac{\pi n} L \right)^2 \bar{\phi}_n^2 \\
  &&+ \frac{V}{\pi}  \frac{\partial }{\partial J_0} \left[\ln
     \left(\frac{u}{VK}\right)\right] \sqrt{\frac{2K}L} \sum_{n=1}^{+\infty}
   [1+(-1)^n] \frac{\pi n} L \bar{\phi}_n.  \nonumber
\end{eqnarray}
The first line gives back the contribution calculated in
App.~\ref{app:correlation}. The second gives the contribution
coming from the edge potential. The necessary Matsubara correlator is
\begin{eqnarray}
\label{eq:chi}
 &&  \left\{\frac{V}{\pi}  \frac{\partial }{\partial J_0} \left[\ln
     \left(\frac{u}{VK}\right)\right]\right\}^2 \frac{2K}L
  \sum_{n=1}^{+\infty} \left(\frac{\pi n}L\right)^2[1+(-1)^n]^2\times
  \nonumber \\
 && \langle T_\tau \bar{\phi}_n(\tau) \bar{\phi}_n(0)\rangle.
\end{eqnarray}
After taking the Fourier transform and making the analytic
continuation, one finds
\begin{eqnarray}
&&\mathrm{Im}\chi_{\mathrm{edge}}(\omega+i0_+)=\left\{\frac{2 V}{\pi}  \frac{\partial }{\partial J_0} \left[\ln
     \left(\frac{u}{VK}\right)\right]\right\}^2\times  \\ && \frac{\pi K}L
  \sum_{n=1}^{+\infty} \sum_{n=1}^{+\infty} \frac{2n\pi}{L} \pi
  \left[\delta\left(\omega -\frac{2\pi n u}{L}\right)
  -\delta\left(\omega + \frac{2\pi n u}{L}\right)\right] \nonumber
\label{eq:imchiapp}
\end{eqnarray}
In the limit of $L\to +\infty$, we end up with
\begin{eqnarray}
  \label{eq:imchi-edge}
  \mathrm{Im}\chi_{\mathrm{edge}}(\omega+i0_+)=\left\{\frac{2 V}{u}  \frac{\partial }{\partial J_0} \left[\ln
     \left(\frac{u}{VK}\right)\right]\right\}^2 \frac{\omega K}{2\pi},
  \nonumber \\
\end{eqnarray}
yielding the edge contribution~(\ref{eq:edge-power}), to be added
to the bulk contribution.

\section{Friedel oscillations}
\label{app:friedel_oscillations}

We consider a Bose-Hubbard chain of $M$ sites with open boundary
conditions. Its Hamiltonian is
\begin{eqnarray}
  \label{eq:bose-hubbard-obc}
  H=-J_0 \sum_{l=1}^{M-1} (b^\dagger_l b_{l+1} + b^\dagger_{l+1} b_l) +
  \frac U 2 \sum_{l=1}^M n_l (n_l-1),
\end{eqnarray}
We  introduce the fictitious sites $0$ and
$M+1$ to write
\begin{eqnarray}
  \label{eq:bose-hubbard-obc2}
  H=-J_0 \sum_{l=0}^{M} (b^\dagger_l b_{l+1} + b^\dagger_{l+1} b_l) +
  U \sum_{j=1}^M n_l (n_l-1),
\end{eqnarray}
and $b_0=b_{M+1}=0$.
The bosonized Hamiltonian reads
\begin{widetext}
\begin{eqnarray}
  \label{eq:hamiltonian}
  H=\int_0^{L} \frac{dx}{2\pi} \left[u K (\pi \Pi)^2 + \frac u K (\partial_x
    \phi)^2 \right] - \frac{V_1}{\pi} \partial_x \phi(0)  -
  \frac{V_2}{\pi} \partial_x \phi(L),
\end{eqnarray}
\end{widetext}
with $L=(M+1)a$ and we have included some forward scattering potentials $V_1,V_2$
at the edges. Our original boson Hamiltonian is symmetric under the
reflection  $b_l \to b_{M+1-l}$. Using the bosonized expressions of the boson
annihilation operators\cite{haldane_bosons,giamarchi_book_1d}, we find
that under reflection
\begin{eqnarray}
  \label{eq:parity}
  P \phi(x) P^\dagger = - \phi(L-x) -\pi N, \\
  P \Pi(x) P^\dagger = -\Pi(L-x),
\end{eqnarray}
so that $V_1=V_2$.
The boundary conditions are derived in the fermion case from
consideration of the non-interacting
limit\cite{fabrizio_open_electron_gas}.
In the boson case, we have to consider the expression of the density:
\begin{eqnarray}
  \label{eq:density}
  \rho(x)=\rho_0 - \frac 1 \pi \partial_x \phi + A \cos (2\phi(x) - 2\pi
  \rho_0 x),
\end{eqnarray}
which implies through the continuity equation that $j=\partial_t \phi/\pi$.
Since no current can leak through the edges of the system, we must
have $\partial_t \phi(0)= \partial_t \phi(L)=0$. So we must impose
the Dirichlet boundary conditions
$\phi(0,t)=\varphi_0$ and $\phi(L,t)=\varphi_1$. Moreover, since the
number of particles in the system is integer, by
integrating~(\ref{eq:density}) we find that
$(\varphi_1-\varphi_0)\pi$ is an integer. We can choose
for instance $\varphi_0=0$, $\varphi_1=-\pi N$, where  $N$ is the number
of particles added to the initial number of particles in the ground
state $N_{GS}$ with
$\rho_0=N_{GS}/L$.
We note that $\partial_x \phi$ can still be non-vanishing as an
operator, so we can \textit{a priori} have edge scattering potentials
$V_1$ and $V_2$ in (\ref{eq:hamiltonian}).

Now, we introduce the Fourier decomposition
\begin{eqnarray}
  \label{eq:phi-decomp}
  \phi(x)&=&-\frac{\pi N x}{L} + \sum_{n=1}^{+\infty}
  \sqrt{\frac{2}{L}} \sin\left(\frac{n\pi x}{L}\right) \phi_n
  e^{-n \frac\epsilon 2}, \\
 \label{eq:Pi-decomp}
\Pi(x)&=&\sum_{n=1}^{+\infty}  \sqrt{\frac{2}{L}} \sin\left(\frac{n\pi x}{L}\right) \Pi_n
  e^{-n \frac\epsilon 2},
\end{eqnarray}
which allows us to rewrite
\begin{eqnarray}
  \label{eq:ham-decomp}
  H&=&\frac{u L}{2\pi K} \left(\frac{\pi N} {L} \right)^2 + \frac{\pi
    N} {L} \frac{V_1+V_2} \pi \nonumber \\
   & & + \frac u 2 \sum_{n=1}^{+\infty} \pi K \Pi_n^2 + \frac 1 {\pi K}
   \left(\frac{\pi n}{L}\right)^2 \phi_n^2 \nonumber \\
   & & - \sum_{n=1}^{+\infty} \sqrt{\frac{2}{L}}  \frac{n}{L}
   [V_1 + (-)^n V_2] \phi_n.
\end{eqnarray}
Until now, we have made no assumption concerning the symmetry of our
bosonized Hamiltonian under parity. Using the Fourier
expansion~(\ref{eq:phi-decomp}), we can show that under a parity
transformation, $P\phi_n P^\dagger = (-1)^n \phi_n$. In the
Hamiltonian (\ref{eq:ham-decomp}) $V_1$ and $V_2$ are exchanged by the
parity transformation. So we recover $V_1=V_2$ for a parity invariant Hamiltonian.
To find the ground state, we have to minimize the first line with
respect to $N$ and determine
the shift of oscillators imposed by the edge potentials.
The minimization with respect to $N$ yields
\begin{eqnarray}
  \label{eq:optimal-n}
  N=E\left(\frac 1 2 - \frac {K}{\pi u} (V_1+V_2)\right),
\end{eqnarray}
while the shift of oscillators is
\begin{eqnarray}
  \label{eq:optim-phi-n}
  \langle \phi_n \rangle = \frac{K}{\pi u} \sqrt{2 L} \frac{V_1 +
    (-)^n V_2} n.
\end{eqnarray}

The expectation value of $\phi(x)$ in the ground state is then
\begin{eqnarray}
  \label{eq:average-phi}
&&   \langle \phi(x) \rangle = -\frac{\pi x} {L} E\left(\frac 1 2 -
    \frac {K}{\pi u} (V_1+V_2)\right) \nonumber \\ && + \frac{2 K}{\pi u}
  \sum_{n=1}^{+\infty} \frac{V_1 + (-)^n V_2} n  \sin\left(\frac{n\pi x}{L}\right)
  e^{-n \frac\epsilon 2}.
\end{eqnarray}

We thus have
\begin{eqnarray}\label{eq:avg-phi-full}
  \langle \phi(x) \rangle = \frac{2 K}{\pi u} \left[V_1 \arctan
    \left(\frac{\sin\left(\frac{\pi x}{L}\right)} {e^{\epsilon/2} -
        \cos\left(\frac{\pi x}{L}\right) }  \right) \right. \nonumber \\ -\left. V_2  \arctan
    \left(\frac{\sin\left(\frac{\pi x}{L}\right)} {e^{\epsilon/2} +
        \cos\left(\frac{\pi x}{L}\right) }  \right) \right] \nonumber
  \\ -\frac{\pi x} {L} E\left(\frac 1 2 -
    \frac {K}{\pi u} (V_1+V_2)\right).
\end{eqnarray}
Taking the limit of $\epsilon \to 0$, for $x$ far enough from an edge,
we find the simplified expression,
\begin{eqnarray}\label{eq:avg-phi}
  \langle \phi(x) \rangle = \frac{K V_1} u - \frac {K(V_1 + V_2)} u
  \frac{x}{L}   -\frac{\pi x} {L} E\left(\frac 1 2 -
    \frac {K}{\pi u} (V_1+V_2)\right),\nonumber \\
\end{eqnarray}
which is a periodic function of
$K(V_1+V_2)/(\pi u)$ of period $1$. So we can restrict ourselves to
$|K(V_1+V_2)/(\pi u)|<1/2$ and drop the integer part in Eq.~(\ref{eq:avg-phi}).
Using Luttinger liquid theory and the expression~(\ref{eq:density}),
we derive
\begin{widetext}
\begin{eqnarray}
  \label{eq:average-rho}
&&  \langle \rho(x)\rangle = \frac 1 {L} \left[ N_{tot.} +
    \frac{K(V_1+V_2)}{\pi u}\right]  \\ && + A \cos\left[\frac{2K V_1} u -
    \frac{2 \pi x}{L} \left( N_{tot.} +
    \frac{K(V_1+V_2)}{\pi u}\right) \right] \left(\frac{\pi
    a}{L \sin\left(\frac{\pi x}{L}\right)}\right)^K, \nonumber
\end{eqnarray}
\end{widetext}
 and we see that far from the edges, the Friedel oscillations  behaves as if the number of particles was
 $N'_{tot.} =   N_{tot.} +
    \frac{K(V_1+V_2)}{\pi u} $. The expression~(\ref{eq:average-rho})
    applies only when $\alpha \ll x$ and $\alpha \ll L-x$. It
    corresponds to the effective Dirichlet boundary conditions $ \phi(0) =
    \frac{K V_1} u$ and $\phi(L)=-\frac{K V_2} u$ that result from the
    phase shift on $\phi(x)$ imposed by the edge potentials. When $0<x<\alpha
    =\epsilon L$, we cannot take the limit $\epsilon \to 0$ in
    Eq.~(\ref{eq:avg-phi-full}). There, $\langle \phi(x) \rangle
    =O(x/\alpha) \to 0$, ensuring that the original Dirichlet boundary
    conditions are satisfied.

\section{Luther-Emery limit}\label{app:luther-emery}

Let's consider the case of the Fermi-Hubbard model in the Mott
insulating phase.
When looking at the bosonized expression of the operator $\mathcal{O}_f$ Eq.~(\ref{eq:bosonized-o-fermions}), we can rescale the fields,  $\Pi_\rho\rightarrow \Pi_\rho/\sqrt{K_\rho}$ and $\phi_\rho\rightarrow \sqrt{K_\rho} \phi_\rho$, such that the operator $\mathcal{O}_{\mathrm{f},\rho}$  becomes:

\begin{eqnarray}
\label{eq:bosonized-o-fermions-rescaled}
\mathcal{O}_{\mathrm{f},\rho}=2 a \sin (k_F a) \int \frac{dx}{2\pi}
\left[\frac{(\pi \Pi_\rho)^2}{K_\rho} + K_\rho (\partial_x \phi_\rho)^2\right].
\end{eqnarray}
The bosonized Hamiltonian for the fermions can also be written in
terms of the rescaled fields and at the Luther-Emery
point\cite{luther_exact}, $K_\rho=\frac 1 2$, it becomes
\begin{equation}
H_\rho=\int \frac{dx}{2\pi} u_\rho \left[ (\pi \Pi_\rho)^2 +
(\partial_x \phi_\rho)^2 \right] - \frac{2
	g_3}{(2\pi\alpha)^2} \int dx \cos 2\phi_\rho.
\end{equation}
That Hamiltonian is rewritten by introducing the pseudofermions
\begin{eqnarray}
\Psi_R(x)= \frac{e^{i(\theta(x)-\phi(x))}}{\sqrt{2 \pi \alpha}},\\
\Psi_L(x)= \frac{e^{i(\theta(x)+\phi(x))}}{\sqrt{2 \pi \alpha}},
\end{eqnarray}
in the form of a gapped non-interacting Hamiltonian
\begin{eqnarray}
H_\rho &=& -iu_\rho \int dx \left( \Psi_R^\dagger \partial_x \Psi_R-\Psi_L \partial_x \Psi_L \right) \\ \nonumber
&-&\Delta \left(\Psi_R^\dagger \Psi_L+\Psi_L^\dagger \Psi_R \right)
\label{eq:luth_emery}
\end{eqnarray}
where we called $\Delta=\frac{2g_3}{4\pi \alpha}$. In terms of the pseudofermions we can rewrite the operator $\mathcal{O}_\rho$ as
\begin{eqnarray}
\label{eq:le_kin}
\mathcal{O}_\rho&=&O_1+O_2 \\
O_1&=&-\frac{5a}{2} i\sin (k_F a) \int dx \left( \Psi_R^\dagger \partial_x \Psi_R-\Psi_L \partial_x \Psi_L\right)\\
O_2&=&3 \pi a \sin(k_F a) \int dx \rho_R\rho_L,
\end{eqnarray}
where $\rho_{L,R}=\Psi^\dagger_{L,R} \Psi_{L,R}$ is the density
operator of the $L,R$ fermions.  We have to evaluate the Matsubara correlator
\begin{equation}
\label{eq:corr}
\chi(\tau)=\sum_{i=1,2}\sum_{j=1,2} \chi_{ij}(\tau)
\end{equation}
with $\chi_{ij}(\tau)=\langle T_\tau O_i(\tau)O_j(0)\rangle$. This correlator can be expressed in terms of the creation and annihilation operators through the representation $\Psi_\nu=\frac{1}{\sqrt{L}}\sum_k e^{i k x}  c_{k,\nu}$, in terms of which the Hamiltonian (\ref{eq:luth_emery}) is written as:
\begin{equation}
H_{\rho}=\sum_k u_\rho k (c^\dagger_{kR}c_{kR}-c^\dagger_{kL}c_{kL})-\Delta (c^\dagger_{kR}c_{kL}+H.c.).
\end{equation}
This Hamiltonian can be diagonalized by standard Bogoliubov transformations and expressed in the form:
\begin{equation}
H'_{\rho}=\sum_k E_k (c^\dagger_{k+}c_{k+}-c^\dagger_{k-}c_{k-}),
\end{equation}
with $E_k=\sqrt{(u_\rho k)^2+\Delta^2}$ and  $c_{kR}=\cos \varphi_k c_{k+}-\sin \varphi_k c_{k-}$, $c_{kL}=\sin \varphi_k c_{k+}+\cos \varphi_k c_{k-}$.
Then the calculations of the correlators (\ref{eq:corr}) proceeds by applying Wick's theorem once the single particle Green's function are known:
\begin{eqnarray}
&& \langle T_\tau
c_{kR(L)}(\tau)c^\dagger_{kR(L)}(0)\rangle=\frac{1}{2} \lbrack
\text{sign}(\tau)\pm \frac{u_\rho k}{E(k)}\rbrack e^{-|\tau| E(k)} \nonumber \\
&&\langle T_\tau c_{kR(L)}(\tau)c^\dagger_{kL(R)}(0)\rangle=\frac{\Delta}{2 E(k)} e^{-|\tau| E(k)}.
\end{eqnarray}
The results for the correlators are:
\begin{widetext}
\begin{eqnarray}
&&\chi_{11}(\tau)=\frac{25}{4} a^2 \sin^2(k_F a) \Delta^2\sum_k \frac{k^2}{E(k)^2}e^{-2|\tau| E(k)}\nonumber \\
&&\chi_{22}(\tau)=\frac{9\pi^2a^2 \sin(k_F
	a)^2}{L^2}\sum_{k_1,\ldots,k_4} \delta_{k_1+k_2,k_3+k_4}e^{-|\tau| \sum_{j=1}^4E(k_j)}\nonumber\\
&&\times  \frac{1}{16}
\left[ \left( \mathrm{sign}(\tau)+\frac{u_{k_4}}{E(k_4)}\right)\left(\mathrm{sign}(\tau)+\frac{u_{k_3}}{E(k_3)}\right)-\frac{\Delta^2}{E(k_3)E(k_4)}\right] \nonumber \\
&&\times
\left[ \left( \mathrm{sign}(-\tau)+\frac{u_{k_1}}{E(k_1)}\right)\left( \mathrm{sign}(-\tau)+\frac{u_{k_2}}{E(k_2)}\right)-\frac{\Delta^2}{E(k_1)E(k_2)}\right] \nonumber \\
&&\chi_{12}(\tau)=\chi_{21}(\tau)=0.
\end{eqnarray}
\end{widetext}
The correlator $\chi_{22}$ can be simplified close to the threshold where an expansion up to order $O(k^2)$ can be performed such that
$\chi_{22}(\tau)\sim \left(\frac{\Delta}{|\tau|}\right)^{7/2} e^{-4|\tau|\Delta}$.
In the complex frequency plane the correlator $\chi_{ij}(i\omega)=\int_{-\infty}^{\infty} d \tau \langle T_\tau O_i(\tau)O_j(0)\rangle$ can be analytically extended to evaluate the imaginary part. The result of the calculation gives:
\begin{eqnarray}
\mathrm{Im} \chi_{11}(i\omega \rightarrow
\omega+i0)=\frac{25a^2\sin(k_F a)^2}{2}L \frac{\Delta^2}{\omega
	u^3}\sqrt{\frac{\omega^2}{4}-\Delta^2}, \nonumber \\
\end{eqnarray}
which shows a threshold at $\omega=2\Delta$ while $\mathrm{Im} \chi_{22}$  has
a threshold at $\omega=4\Delta$. The resulting absorbed power in the Luther-Emery limit is
\begin{eqnarray}
\label{eq:le-power-app}
\mathcal{P}_{LE}=\frac{L\Delta^2} u \left(\frac{5a \delta J} {2u} \sin(k_F
a)\right)^2 \sqrt{\left(\frac\omega 2\right)^2 -\Delta^2},
\end{eqnarray}
for $2\Delta <\omega <4\Delta$.
This analysis can be extended away from the Luther-Emery point to any
value of $K=K_\rho$ by using the form factor expansion for the
sine-Gordon
model~\cite{karowski_ff,babujian99_ff_sg_1,babujian02_ff_sg_2,essler04_condmat_exact_review},
as detailed in the App.\ref{app:form_factor} (see also \cite{iucci_absorption}).

\section{The form factor approach}
\label{app:form_factor}

In the present Appendix, we want to extend the results derived using
the Luther-Emery limit to any value of $K_\rho$. For $K_\rho>1/2$ the
excitations are massive solitons and antisolitons of mass $M_s$, while
for $K<1/2$ we also have breathers of mass $M_n=2M_s \sin\left( n
  \frac \pi 2 \frac{K}{1-K}\right)$ with $1\le n < \frac{1}{K_\rho}-1$
integer.
Working in the vicinity of
the Luther-Emery point the low-energy Hamiltonian is:

\begin{eqnarray}
&&H=-i\frac{u_\rho}{2} \left( 2K_\rho+\frac{1}{2K_\rho}\right) \int dx \left(
   \Psi_R^\dagger \partial_x \Psi_R-\Psi_L \partial_x \Psi_L \right)
  \nonumber \\
  && -\frac{g}{2\pi \alpha} \left(\Psi_R^\dagger \Psi_L+\Psi_L^\dagger
     \Psi_R \right) \nonumber \\  && -\pi u_\rho \left( 2K_\rho-\frac{1}{2K_\rho}\right)\int dx \hat{\rho}_R \hat{\rho}_L,
\label{eq:away_luth_emery}
\end{eqnarray}
where the operator $\hat{\rho}_a=\Psi_a^\dagger \Psi_a$, $a=L,R$. It
is the Hamiltonian of a massive Thirring
model\cite{thacker_bethe_review}:
\begin{eqnarray}
&&H=-i\bar{v} \int dx \left( \Psi_R^\dagger \partial_x
   \Psi_R-\Psi_L \partial_x \Psi_L \right) \nonumber \\
&&  -M \int dx \left(\Psi_R^\dagger \Psi_L+\Psi_L^\dagger \Psi_R \right)\nonumber
  \\
&&  - \bar{g}\int dx \hat{\rho}_R \hat{\rho}_L,
\label{eq:thirring}
\end{eqnarray}
where  $\bar{v}=\frac{u}{2} \left( 2K+\frac{1}{2K}\right)$,
$M=\frac{g}{2\pi \alpha}$ and $\bar{g}=\pi u_\rho \left( 2K-\frac{1}{2K}\right)$.

The kinetic energy operator is related to the component $T^{11}$ of the
momentum-energy tensor (see \cite{itzykson-zuber} p. 143), so that
\begin{eqnarray}
  \frac{u_\rho}{2a \sin (k_F
    a)}\mathcal{O}&=&\frac{2K_\rho}{2K_\rho^2+1} \int dx T^{11}(x) \\ && + 2 \pi u_\rho \frac{K_\rho^2
    -1}{K_\rho^2+1} \int dx \hat{\rho}_R \hat{\rho}_L. \nonumber
\end{eqnarray}
Since $\hat{\rho}_R \hat{\rho}_L=\hat{\psi}^\dagger_R \hat{\psi}^\dagger_L \hat{\psi}_R \hat{\psi}_L$,
that operator can only have matrix elements between the ground state
of the massive Thirring model and a state containing two solitons and
two antisolitons,
{i.e.}, a state with energy at least $4M_s$. So that term
will be not contribute  for frequencies $\omega<4 M_s$, and we will
have
\begin{eqnarray}
  \mathrm{Im} \chi_\rho(\omega) = L
  \left(\frac{4 K_\rho a \sin (k_F a)}{u_\rho (K_\rho^2+1)}\right)^2 \mathrm{Im} \chi^{s\bar
    s}(\omega),
\end{eqnarray}
where the contribution $ \mathrm{Im} \chi^{s\bar
    s}(\omega)$ of the $T^{11}$ component
of the momentum-energy tensor is
obtained from the form-factor
expansion\cite{babujian99_ff_sg_1,karowski_ff,babujian02_ff_sg_2}.
For the lowest excited state formed of a  single
soliton-antisoliton pair, we have
\begin{widetext}
\begin{eqnarray}
&&  \mathrm{Im} \chi^{s\bar s}(\omega) = 2\pi  \int \frac{d\theta
    d\bar\theta}{(2\pi)^2} |\langle 0
  |T^{11}|\theta,\bar\theta\rangle_{s\bar s}|^2 \delta\left(\frac {M_s} {u_\rho} (\sinh
  \theta + \sinh \bar \theta)\right)  \nonumber \\
&&  \times \delta\left(\omega - M_s  (\cosh
    \theta + \cosh \bar \theta)\right),
\end{eqnarray}
\end{widetext}
where according to \cite{babujian02_ff_sg_2}, the form factor of the
energy momentum tensor is:
\begin{eqnarray}
&&  \langle 0
  |T^{11}|\theta_1, \theta_2\rangle_{s\bar s} =-2 i \frac {M^2} u
  \cosh^2\left(\frac{\theta_1+\theta_2} 2 \right) \times \\ && \sinh
  \left(\frac{\theta_1-\theta_2} 2 \right) \frac{F_+(\theta_1
    -\theta_2)}{\nu}, \nonumber \\
&& F_+(\theta)=\frac{i \cosh (\theta/2)}{\sinh \left(\frac{i\pi
       -\theta}{2\nu}\right)} F_{\mathrm{min}}(\theta) \\
&& F_{\mathrm{min}}(\theta)=\exp\left[\int_0^{+\infty} \frac {dt}{t}
  \frac{\sinh \frac 1 2 (1-\nu) t}{\sinh \frac {\nu t} 2 \cosh \frac t
    2 } \frac {1 -\cosh \left(1-\frac \theta {i\pi}\right) t}{2 \sinh
    t}\right]  \nonumber \\
\end{eqnarray}
Here $\nu=K_\rho/(1-K_\rho)$ for the Fermi-Hubbard model and $\nu=K/(2-K)$ for the Bose-Hubard model. When $\theta\rightarrow 0$, $F_+(\theta)\rightarrow 1$, giving for
$\omega\rightarrow 2M^+$,

\begin{eqnarray}
\mathrm{Im}\chi^{s\bar{s}} (i\omega \rightarrow \omega+i0) =   \frac{4M_s^3}{\pi \omega u_\rho \nu^2}\sqrt{\left(\frac{\omega}{2M_s}\right)^2-1}.
\end{eqnarray}
Now we have a threshold at twice the mass of the soliton. A similar
threshold behavior was also obtained in the case of modulation of a
weak optical lattice\cite{iucci_shake_bosons_theory}. Technically,
this can be understood as follows. We can always substract an operator
proportional to the Hamiltonian~(\ref{eq:thirring}) from the operator
$\mathcal{O}$. So we would obtain an equivalent result if $T^{11}$ was
replaced by a term proportional to $\Psi^\dagger_R \Psi_L+
\Psi^\dagger_L \Psi_R$, which is precisely the perturbing term
in\cite{iucci_shake_bosons_theory}.

%\bibliography{totphys-cleaned,modulationnew,refsoutlook}

\end{document}